\documentclass[prb,twocolumn,a4paper,showpacs,superscriptaddress]{revtex4}
\usepackage{amsfonts,amssymb}
\usepackage[intlimits]{amsmath}
\usepackage[english]{babel}
\usepackage{epsfig}

\begin{document}

\title{An exact Coulomb cutoff technique for supercell calculations}

\author{Carlo A. Rozzi}
\affiliation{Institut f\"ur theoretische Physik, Freie Universit\"at Berlin,
Arnimallee 14, D-14195 Berlin, Germany}
\affiliation{European Theoretical Spectroscopy Facility (ETSF)}

\author{Daniele Varsano}
\affiliation{Departamento de F\'{\i}sica de Materiales, Facultad de Ciencias Qu\'{\i}micas, UPV/EHU,
Centro Mixto CSIC-UPV/EHU and Donostia International Physics Center, E-20018 San Sebasti\'an, Spain}
\affiliation{European Theoretical Spectroscopy Facility (ETSF)}

\author{Andrea Marini}
\affiliation{Dipartimento di Fisica, Universit\`a ``Tor Vergata", Roma, Italy}
\affiliation{European Theoretical Spectroscopy Facility (ETSF)}

\author{Eberhard K. U. Gross}
\affiliation{Institut f\"ur theoretische Physik, Freie Universit\"at
Berlin, Arnimallee 14, D-14195 Berlin, Germany}
\affiliation{European Theoretical Spectroscopy Facility (ETSF)}

\author{Angel Rubio}
\affiliation{Institut f\"ur theoretische Physik, Freie Universit\"at
Berlin, Arnimallee 14, D-14195 Berlin, Germany}
\affiliation{Departamento de F\'{\i}sica de Materiales, Facultad de Ciencias Qu\'{\i}micas, UPV/EHU,
Centro Mixto CSIC-UPV/EHU and Donostia International Physics Center, E-20018 San Sebasti\'an, Spain}
\affiliation{European Theoretical Spectroscopy Facility (ETSF)}


\begin{abstract}
We present a new reciprocal space analytical method to cutoff the long range
interactions in supercell calculations for systems that are infinite and
periodic in 1 or 2 dimensions, extending previous works for finite systems. The
proposed cutoffs are functions in Fourier space, that are used as a
multiplicative factor to screen the bare Coulomb interaction. The functions are
analytic everywhere but in a sub-domain of the Fourier space that depends on the
periodic dimensionality. We show that the divergences that lead to the
non-analytical behaviour can be exactly cancelled when both the ionic and the
Hartree potential are properly screened. This  technique is exact, fast, and
very easy to implement in already existing supercell codes. To illustrate  the
performance of the new scheme, we apply it to the case of the Coulomb 
interaction in systems with reduced periodicity (as one-dimensional chains and
layers). For those test cases we address the impact of the cutoff in  different
relevant quantities for ground and excited state properties, namely: the
convergence of the ground state properties, the static polarisability of the
system, the quasiparticle corrections in the GW scheme and in the binding energy
of the excitonic states in the Bethe-Salpeter equation. The results are very
promising.
\end{abstract}


\date{\today}
\maketitle

\section{Introduction}

%
%
Plane waves expansions with periodic boundary conditions have been proven to be
a very effective way to exploit the translational symmetry of infinite crystal
solids, in order to calculate the properties of the bulk, by performing the
simulations in one of its primitive cells only~\cite{pw}. The use of plane waves
is motivated by several facts. First, the translational symmetry of the
potentials involved in the calculations is naturally and easily accounted for in
reciprocal space, through the Fourier expansion. Second, very efficient and fast
algorithms exist (like FFTW\cite{fftw}) that allow us to calculate the Fourier
transforms very efficiently. Third the expansion in plane waves is exact, since
they form a complete set, and it is only limited in practice by one parameter,
namely, the maximum value of the momentum, that determines the size of the
chosen set. Fourth, in many cases, the use of Born-von Karm\'an periodic
boundary conditions supplies a conceptually easy (though artificial) way to get
rid of the dependence of the properties of a specific sample on its surface and
shape, allowing us to concentrate on the bulk properties of the system in the
thermodynamic limit.\cite{Lebowitz69}

However, mainly in the last decade, increasing interest has been developed in
systems at the nano-scale, like tubes, wires, quantum-dots, biomolecules, etc.,
whose physical dimensionality is, for all practical purposes, less than
three.\cite{nano} These systems are still 3-dimensional (3D), but their quantum
properties are those of a confined system in one or more directions, and those
of a periodic extended system in the remaining directions. Other classes of
systems with the same kind of reduced periodicity are the classes of the
polymers, and of the solids with defects.

Throughout this paper we call nD-periodic a 3D object, that can be considered
infinite and periodic in $n$ dimensions, being finite in the remaining $3-n$
dimensions. In order to simulate this kind of systems, a commonly adopted
approach is the supercell approximation.\cite{pw}

In the supercell approximation the physical system is treated as a fully
3D-periodic one, but a new unit cell (the supercell) is built in such a way that
some extra empty space separates the periodic replica along the direction(s) in
which the system is to be considered as finite. This method makes possible to
retain all the advantages of plane waves expansions and periodic boundary
conditions. Yet the use of a supercell to simulate objects that are not infinite
and periodic in all the directions, leads to artifacts, even if a very large
portion of vacuum is interposed between the replica of the system in the
non-periodic dimensions.

In fact, the straightforward application of the supercell method generates in
any case fake images of the original system, that can mutually interact in
several ways, affecting the results of the simulation. It is well known that the
response function of an overall neutral solid of molecules is not equal, in
general, to the response of the isolated molecule, and converges very slowly to
it, when the amount of vacuum in the supercell is
progressively increased.\cite{Onida02,ijm}

For instance, the presence of higher order multipoles can make undesired images
interact via the long range part of the Coulomb potential. In the dynamic
regime, multipoles are always generated by the oscillations of the charge
density even in systems whose unit cell does not carry any multipole in its
ground state. This is the case, for example, when we investigate the response of
a system in presence of an external oscillating electric field.

Things go worse when the unit cell carries a net charge, since the total charge
of the infinite system represented by the supercell is actually infinite, while
the charges at the surfaces of a finite, though very large system always
generate a finite polarisation field. This situation is usually normalised in
the calculation by the introduction of a suitable compensating positive
background charge.

Another common situation in which the electrostatics is known to modify the
ground state properties of the system occurs when a layered system is studied,
and an infinite array of planes is considered instead of a single slab, being in
fact equivalent to an effective chain of capacitors.\cite{Wood04}

These issues become particularly evident in all the approaches that imply the
calculation of non-local operators or response functions, because, in these
cases, two supercells may effectively interact even if their charge densities do
not overlap at all. This is the case, for example, of the many-body perturbation
theory calculations (MBPT), and, in particular, of the self-energy calculations
at the GW level.\cite{GW,Onida02}

However we are usually still interested in the dispersion relations of the
elementary excitations of the system along its periodic directions, and those
are ideally dealt with using a plane waves approach. Therefore, the ideal path
to keep the advantages of the supercell formulation in plane waves, and to gain
a description of systems with reduced periodicity free of spurious effects is to
develop a technique to cut the Coulomb interaction off out of a desired region.
This problem is not new and has been addressed now for a very long time and in
different fields (condensed matter, classical fields, astrophysics, biology,
etc). Several different approaches have been proposed in the past to solve it,
however a complete review of them is beyond the scope of this paper.  The aim
of the present work is to focus on the widely used supercell schemes to show how
the image interaction influences both the electronic ground state properties and
the dynamical screening in the excited state of 0D-, 1D-, 2D-periodic systems,
and to propose an exact method to avoid the undesired interaction of the
replicas in the non-periodic directions.

The paper is organised as follows: in Sec.~\ref{sect-3D} the basics of the plane
wave method for solids are reviewed, in Sec.~\ref{sect-nD} the new method is
outlined, in Sec.~\ref{sect-singular} the treatment of the singularities is
explained, in Sec.~\ref{sect-results} some applications of the proposed
technique are discussed.

\section{The 3D-periodic case}\label{sect-3D}

The main problem of electrostatics we are facing here
can be reduced to the problem of finding
solutions to the Poisson equation for a given charge distribution $n(\mathbf{r})$,
and given boundary conditions
\begin{equation}\label{PoissonEq}
  \nabla^2V(\mathbf{r}) = -4\pi n(\mathbf{r}).
\end{equation}
In a finite system the potential is usually required to be zero at
infinity. In a periodic system this condition is meaningless, since the system
itself extends to infinity. Nevertheless the general solution of Eq.
(\ref{PoissonEq}) in both cases is known in the form of the convolution
\begin{equation}\label{defHartree}
  V(\mathbf{r})=
  \iiint_{space}
  \frac{n(\mathbf{r'})}{|\mathbf{r}-\mathbf{r'}|}
  \mathrm{d}^3\mathbf{r}',
\end{equation}
that it is referred from now on as the Hartree potential.

It might seem that the most immediate way to build the solution potential for a
given charge distribution is to compute the integral in real space, but problems
immediately arise for infinite systems. In fact, the density can be reduced to
an infinite sum over delta charge distributions
$q\delta(\mathbf{r}-\mathbf{r'})$, 
\begin{equation}\label{1DCoulombSum}
  V(\mathbf{r})=
  \sum_{\mathbf{n}}
  \frac{q}{|\mathbf{r}-\mathbf{L}_{\mathbf{n}}|},
\end{equation}
and the integral in Eq. (\ref{defHartree}) becomes an infinite sum as well, but
this sum is in general only conditionally, and not absolutely
convergent.\cite{Makov95} The sum of Eq.~(\ref{1DCoulombSum}) a potential that
is determined up to a constant for a neutral cell with zero dipole moment, while
the corresponding sum for the electric field is absolutely convergent. A neutral
cell with a non null dipole moment, on the opposite, gives a divergent
potential, and an electric field that is determined up to an unknown constant
electric field (the sum for the electric field is conditionally convergent in
this case).

Even if, in principle, the surface terms have to be always taken into account,
in practice they are only relevant when we want to calculate energy differences
between states with different total charge. These terms can be neglected in the
case of a neutral cell whose lowest nonzero multipole is
quadrupole\cite{DeLeeuw80}. As in the present work we are interested in
macroscopic properties of the periodic system, those surface effects are never
considered in the discussion that follows. However this sample-shape effects
play an important role for the analysis of different spectroscopies as, for
example, infrared and nuclear magnetic resonance.

A major source of computational problems is the fact that the sum in Eq.
(\ref{1DCoulombSum}) is very slowly converging when it is summed in real space,
and this fact has historically motivated the need for reciprocal space methods
to calculate it. It was Ewald who first discovered that, by means of an integral
transform, the sum can be split in two terms, and that if one is summed in real,
while the other in reciprocal space, both of them are rapidly
converging.\cite{Ewald21} The point of splitting is determined by an arbitrary
parameter.

Let us now focus on methods of calculating the sum in Eq.~(\ref{1DCoulombSum})
purely based on the reciprocal space.

If we consider a periodic distribution of charges with density $n(\mathbf{r})$ such that $n(\mathbf{r})=n(\mathbf{r}+\mathbf{L_{\mathbf{n}}})$, with $\mathbf{L_{\mathbf{n}}}=\{n_x \mathbf{L}_x,n_y \mathbf{L}_y,n_z \mathbf{L}_z\}$, and $\{n_x,n_y,n_z\}\in\mathbb{Z}$, it turns out that the reciprocal space expression for a potential like
\begin{equation}
  V(\mathbf{r})=
  \iiint_{space}
  n(\mathbf{r'})v(|\mathbf{r}-\mathbf{r'}|)
  \mathrm{d}^3\mathbf{r}',
\end{equation}
in a 3D-periodic system, can be written as
\begin{equation}\label{defHartreeReciproc}
  V(\mathbf{G_{\mathbf{n}}})=
  n(\mathbf{G_\mathbf{n}})v(\mathbf{G}_{\mathbf{n}}),
\end{equation}
where we have used the convolution theorem to transform the real space convolution of the density and the Coulomb potential into the product of their reciprocal space counterparts. Here $\mathbf{G_{\mathbf{n}}}=\{n_x \mathbf{G}_x,n_y \mathbf{G}_y,n_z \mathbf{G}_z\}$ are the multiples of the primitive reciprocal space vectors $\frac{2\pi}{\mathbf{L_{\mathbf{n}}}}$, and $v(\mathbf{G}_{\mathbf{n}})$ is the Fourier transform of the long range interaction $v(\mathbf{r})$, evaluated at the point $\mathbf{G}_{\mathbf{n}}$. For the Coulomb potential it is
\begin{equation}\label{CoulombRecipr}
  v(\mathbf{G_{\mathbf{n}}})= \frac{4\pi}{G_{\mathbf{n}}^2}.
\end{equation}

Fourier transforming expression (\ref{defHartreeReciproc}) back into real space
we have, for a unit cell of volume $\Omega$,
\begin{equation}
  V(\mathbf{r})=\frac{4\pi}{\Omega}
  \sum_{\mathbf{n}\neq 0}\frac{n(\mathbf{G_\mathbf{n}})}{G_{\mathbf{n}}^2}\exp(i \mathbf{G_\mathbf{n}}\cdot \mathbf{r}).
\end{equation}

At the singular point $n_x=n_y=n_z=0$ the potential $V$ is undefined, but, since
the value at $G=0$ corresponds to the average value of $V$ in real space, it can
be chosen to be any number, corresponding to the arbitrariness in the choice of
the static gauge (a constant) for the potential. Observe that the same
expression can be adopted in the case of a charged unit cell, but this time, the
arbitrary choice of $v(G)$ in $G=0$ corresponds to the use of a uniform
background neutralising charge.

\section{Systems with reduced periodicity}\label{sect-nD}
  
It has been shown\cite{Sphor94} that the slab capacitance effect mentioned in
the introduction actually is a problem that cannot be solved by just adding more
vacuum to the supercell. This has initially led to the development of
corrections to Ewald's original method\cite{Yeh99}, and then to rigorous
extensions in 2D and 1D.\cite{Martyna99,Brodka03} The basic idea is to restrict
the sum in reciprocal space to the reciprocal vectors that actually correspond
to the periodic directions of the system. These approach are in general of order
$O(N^2)$\cite{Heyes77,Grzybowski00}, but they have been recently refined to
order $O(N\ln N)$.\cite{Minary02,Minary04} Another class of techniques,
developed so far for finite systems, is based on the expansion of the
interaction into a series of multipoles (fast multipole method).\cite{fmm,Castro03,fmm-perio} With this technique it is possible to
evaluate effective boundary conditions for the Poisson's equations at the cell's
boundary, so that the use of a supercell is not required at all, making it
computationally very efficient for finite\cite{fmm,Castro03} and extended
systems\cite{fmm-perio}. Other known methods, tipically used in molecular
dynamics simulations, are the multipole-correction method\cite{Schultz99}, and
the particle-mesh method\cite{Hockney}, whose review is beyond the scope of the
present work, and we refer the reader to the original works for details.

Differently from what happens for the Ewald sum, the method that we propose to
evaluate the sum in Eq.~(\ref{1DCoulombSum}) entirely relies on the Fourier
space and amounts to screening the unit cell from the undesired effect of (some
of) its periodic images. The basic expression is Eq. (\ref{defHartreeReciproc}),
whose accuracy is only limited by the maximum value $G_N$ of the reciprocal
space vectors in the sum. Since there is no splitting between real and
reciprocal space, no convergence parameters are required.

Our goal is to transform the 3D-periodic Fourier representation of the
Hartree potential of Eq. (\ref{defHartreeReciproc}) into the modified one
\begin{equation}\label{defHartreeReciprocScreen}
  \tilde V(\mathbf{\mathbf{G}_{\mathbf{n}}})=
  \tilde n(\mathbf{G_\mathbf{n}})\tilde v(\mathbf{G}_{\mathbf{n}})
\end{equation}
such that all the interactions among the undesired periodic replica
of the system disappear. The present method is a generalisation of
the method proposed by Jarvis {\em et al.}\cite{Jarvis97} for the
case of a finite system.

In order to build this representation, we want to: 1) define a screening region
$\cal D$ around each charge in the system, out of which there is no Coulomb
interaction; 2) calculate the Fourier transform of the desired effective
interaction $\tilde v(r)$ that equals the Coulomb potential in  $\cal{D}$, and
is $0$ outside $\cal{D}$

\begin{equation}
  \tilde V(r) =
  \begin{cases}
    \frac{1}{r}&    \text{if $r\in\cal{D}$} \\
    0&          \text{if $r\notin\cal{D}$}
  \end{cases}.
\end{equation}
Finally we must 3) modify the density $n(\mathbf{r})$ in such a way that the
effective density is still 3D-periodic, so that the convolution theorem can be
still applied, but densities belonging to undesired images are not close enough
to interact through $\tilde v(r)$.

The choice of the region $\cal{D}$ for step 1) is suggested by symmetry
considerations, and it is a sphere (or radius $R$)
for finite systems, an infinite cylinder (of radius $R$) for
1D-periodic systems, and an infinite slab (of thickness $2R$)
for 2D-periodic systems.

Step 2) means that we have to calculate the modified Fourier integral
  \begin{equation}
    \tilde V(\mathbf{G})=\iiint_{space}
    \tilde v(r)\mathrm{e}^{-i\mathbf{G}\cdot\mathbf{r}}
    \mathrm{d}^3\mathbf{r} =
    \iiint_{\cal D}
    v(r)\mathrm{e}^{-i\mathbf{G}\cdot\mathbf{r}}
    \mathrm{d}^3\mathbf{r}.
  \end{equation}
Still we have to avoid that two neighbouring images interact by taking them far
away enough from each other. Then step 3) means that we have to build a suitable
supercell, and re-define the density in it.

Let us examine first step 2), i.e. the cutoff Coulomb interaction in reciprocal space. We know the expression of the potential when it is cutoff in a sphere.\cite{Jarvis97} It is
\begin{equation}\label{0Dcutoff}
  \tilde v^{0D}(G) = \frac{4\pi}{G^2}\bigl[1-\cos(GR)\bigr].
\end{equation}
The limit $R\to\infty$ converges to the bare Coulomb term  in the sense of a
distribution, while, since $\lim_{G\to 0} \tilde v^{0D}(G)= 2\pi R^2$,  there is
no particular difficulty in the origin. This scheme has been successfully used
in many  applications\cite{Makov95,Jarvis97,Castro03,octopus,Onida95}.

The 1D-periodic case applies to systems with infinite extent in the $x$
direction, and finite in the $y$ and $z$ directions. The effective Coulomb
interaction is then defined in real space  to be $0$ out of a cylinder of radius
$R$ having its axis parallel to the $x$ direction. By performing the Fourier 
transformation we get the following expression for the cutoff coulomb potential
in cylindrical coordinates:\cite{Gradshteyn}
\begin{multline}
  \tilde v^{1D}(G_x,G_{\bot}) = \frac{4\pi}{G^2}
  \bigg[1 + G_{\bot}RJ_1(G_{\bot}R)K_0(|G_x| R) \label{1DCutoff}\\
          - |G_x| RJ_0(G_{\bot}R)K_1(|G_x| R)\bigg],
\end{multline}
where $J$ and $K$ are the ordinary and modified cylindrical Bessel functions,
and $G_{\bot}=\sqrt{G_y^2+G_z^2}$.

It is easy to realise that, since the $K$ functions damp the oscillations of
the $J$ functions very quickly, for all practical purposes this cutoff function
only acts on the very first smaller values of $G$, while the unscreened
$\frac{4\pi}{G^2}$ behaviour is almost unchanged for the larger values.

Unfortunately, while the $J_n(\xi)$ functions have a constant value for
$\xi=0$, and the whole cutoff is well defined for $G_{\bot}=0$, the
$K_0(\xi)$ function diverges logarithmically for $\xi\to0$. Since, on the
other hand, $K_1(\xi)\approx \xi^{-1}$ for small $\xi$,
\begin{equation}
  \tilde v^{1D}(G_x,G_{\bot})
  \sim -\log(G_xR)\quad \textrm{for $G_{\bot}>0$, $G_x\to 0^+$}.
\end{equation}
This means that the limit $\lim_{G\to0^+}v(G)$ does not exist for this cutoff
function, and the whole $G_x=0$ plane is ill-defined. We will come back to the
treatment of the singularities in the next section. We notice that this
logarithmic divergence is the common dependence one would get for the
electrostatic potential of a uniformly charged 1D wire\cite{Jackson}. It is
expected that bringing charge neutrality in place would cancel this divergence
(see below).

The 2D-periodic case, with finite extent in the $z$ direction, is calculated in
a similar manner. The effective Coulomb interaction is defined in real space
to be $0$ out of a slab of thickness $2R$ symmetric with respect to the $xy$ plane. In Cartesian coordinates we get
\begin{align}\label{2DCutoff}
  \tilde v^{2D}(G_{\|},G_z) = \frac{4\pi}{G^2}
  \bigg[1  &+ \mathrm{e}^{-G_{\|} R}\frac{|G_z|}{G_{\|}}\sin(|G_z| R) \nonumber \\
       &- \mathrm{e}^{-G_{\|} R}\cos(|G_z| R)\bigg],
\end{align}
where $G_{\|}=\sqrt{G_x^2+G_y^2}$.

In the limit $R\to\infty$ the unscreened potential $\frac{4\pi}{G^2}$ is
recovered. Similarly to the case of 1D, the limit $G\to0$ does not exist, since
for $G_z=0$, the cutoff has a finite value, while it diverges in the limit
$G_{\|}\to0$
\begin{equation}
  \tilde v^{2D}(G_{\|},G_z)
  \sim \frac{1}{G_{\|}^2}\quad \textrm{for $G_{\|}>0$, $G_z\to 0^+$}.
\end{equation}

So far we haven't committed to a precise value of the cutoff length $R$. This value has to be chosen, for each dimensionality, in such a way that it avoids the interaction of any two neighbour images of the unit cell in the non-periodic dimension.

In order to fix the values of $R$ we must choose the size of the supercell. This
leads us to the step 3) of our procedure. We recall that even once the long
range interaction is cutoff out of some region around each component of the
system, this is not sufficient yet to avoid the interaction among undesired
images. The charge density has to be modified, or, equivalently, the supercell
has to be built in such a way that two neighbouring densities along every non
periodic direction do not interact via the cutoff interaction.

\begin{figure}
    \centering
    \epsfig{file=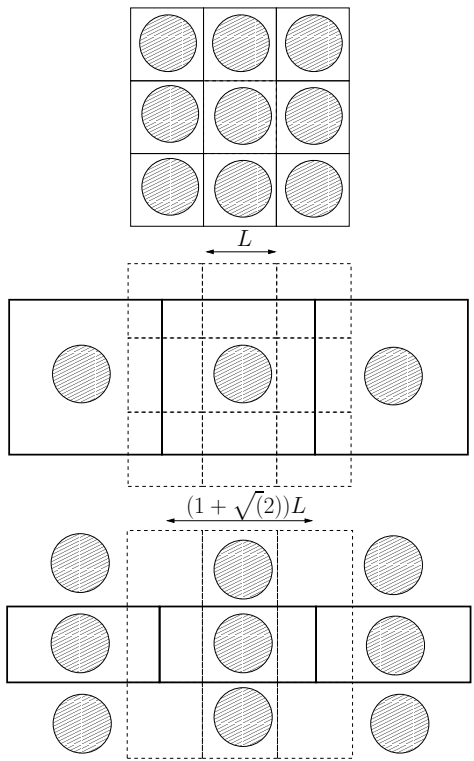,width=0.72\columnwidth}
    \caption{Schematic description
    for  the supercell construction in a 2D system. The upper sketch corresponds to the 2D-periodic case (i.e. a 2D crystal). The middle sketch corresponds to a 0D-periodic system, and the bottom one to a 1D-periodic. In the 0D-periodic case the systems in different cells do not interact, while in the 1D-periodic the chains do not interact, but within each chain the interaction of all of its elements is permitted.\label{fig-pad_all}.}
\end{figure}

It is easy to see how this could happen in the simple case of a 2D square cell
of length $L$: if both $\mathbf{r}$ and $\mathbf{r'}$ belongs to the cell, then
$r,r'\leq L$, and $|\mathbf{r}-\mathbf{r'}|\leq \sqrt{2}L$ (see the schematic
drawing in Fig.~\ref{fig-pad_all}). If a supercell is built that is smaller than
$(1+\sqrt{2})L$, there could be residual interaction, and the cutoff would no
longer lead to the exact removal of the undesired interactions.

Let us call $A_0$ the unit cell of the system we are working on, and ${\cal
A}=\{A_i,i=-\infty,\cdots,\infty\}$ the set of all the cells in the system. If
the system is nD-periodic this set only includes the periodic images of $A_0$
in the $n$ periodic directions. Let us call ${\cal B}$ the set of all the
non-physical images of the system, i.e. those in the non-periodic directions.
Then ${\cal A}\cup{\cal B}=\mathbb{R}^3$. Obviously, if the system is
3D-periodic ${\cal A}=\mathbb{R}^3$, and ${\cal B}\equiv\varnothing$.

In general we want to allow the interaction of the electrons in $A_0$  with the
electrons in all the cells $A_i\in{\cal A}$, but not with those $B_i\in{\cal
B}$. To obtain this we define the supercell $C_0\supseteq A_0$ such that, $\forall i$
\begin{equation}
  \begin{cases}
  \text{if $\mathbf{r}\in A_0$, and $\mathbf{r}'\in A_i$} & \text{then $|\mathbf{r}-\mathbf{r}'|\in C_0$}    \\
  \text{if $\mathbf{r}\in A_0$, and $\mathbf{r}'\in B_i$} & \text{then $|\mathbf{r}-\mathbf{r}'|\notin C_0$}  \\
  \end{cases}
\end{equation}
(see Fig. \ref{fig-pad_all} for a simplified 2D
sketch). The new density $\tilde n(\mathbf{r})$ is such that
\begin{equation}
  \begin{cases}
  \text{if $\mathbf{r}\in A_0$} & \text{then $\tilde n(\mathbf{r})=n(\mathbf{r})$,}    \\
  \text{if $\mathbf{r}\in C_0$, and $\mathbf{r}\notin A_0$} & \text{then $\tilde n(\mathbf{r})=0$.}  \\
  \end{cases}
\end{equation}
The size $L_C$ of the super-cell in the non-periodic directions depends on the
periodic dimensionality of the system. In order to completely avoid any
interaction, even in the case the density of the system is not zero at the cell
border, it has to be
\begin{equation}
  \begin{cases}
  \text{$L_C=(1+\sqrt3)L$} & \text{for finite}      \\
  \text{$L_C=(1+\sqrt2)L$} & \text{for 1D-periodic} \\
  \text{$L_C=2L$}          & \text{for 2D-periodic}
  \end{cases}
\end{equation}
Actually, since the required super-cell is quite large, a compromise between
speed and accuracy can be achieved in the computation, using parallelepiped
super-cell with $L_C = 2L$ for all the cases. This approximation rests on the
fact that the charge density is usually contained in a region smaller than the
cell in the non-periodic directions, so that the spurious interactions are, in
fact, avoided, even with a smaller cell. Therefore, on the basis of this
approximation, we can choose the value of the cutoff length $R$ always as half
the smallest primitive vector in the non-periodic dimension.

\section{Cancellation of the singularities}\label{sect-singular}

The main point in the procedure of eliminating the divergences in all the cases
of interest is to observe that our final goal is usually not to obtain the
expression of the Hartree potential {\em alone}, because all the physical
quantities depend on the total potential, i.e. on the sum of the electronic and
the ionic potential. When this sum is considered we can exploit the fact that
each potential is defined up to an arbitrary additive constant, and choose the
constants consistently for the two potentials. Since we know in advance that the
sum must be finite, we can include, so to speak, all the infinities into these
constants, provided that we find a method to separate out the long range part of
both potentials on the same footing. 

In what follows we show how charge neutrality can be exploited to obtain the
{\it exact cancellations} when operating with the cutoff expression of Sec.
\ref{sect-nD} in Fourier space.

The total potential of the system is built in the following way: we separate out
first short and long range contributions to the ionic potential by adding and
subtracting a Gaussian charge density $n_+(r)=Z\exp(-a^2r^2)$. The potential
generated by this density is  $V_+(r) = Z\frac{\mathrm{erf}(ar)}{r}$. The ionic potential is then written as
\begin{equation}
  V(r) = \Delta V(r) - Z\frac{\mathrm{erf}(ar)}{r},
\end{equation}
where $a$ is chosen so that $\Delta V(r)$ is localised within a sphere
of radius $r_a$, smaller than the cell size.
The expression of the ionic potential in reciprocal space is
\begin{equation}
  V(\mathbf{G}) =
4\pi \left[ \int_0^{+\infty}  \frac{r\sin(Gr)}{G} \Delta V(r)\mathrm{d}r
- \frac{\exp\left(-\frac{G^2}{4a^2}\right)}{G^2} \right]
\end{equation}
which, for $G=0$ gives a finite contribution from the first term,
and a divergent contribution from the second term
\begin{equation}\label{local-infty}
  V(G=0) = 4\pi \int_0^{+\infty} r^2 \Delta V(r)\mathrm{d}r-\infty.
\end{equation}
The first is the contribution of the localised charge, and is easily computed,
since the integrand is zero for $r>r_a$. The second term is cancelled by the
corresponding $G=0$ term in the electronic Hartree potential, due to the charge
neutrality of the system. This trivially solves the problem of the divergences
in 3D-periodic systems.

Now let us consider a 1D-periodic system. The Hartree term alone in
real space is given by
\begin{multline}
  V(x,y,z) = \sum_{G_x}\iint_{\Omega} n(G_x,y',z') \label{HartreeRealRecipr} \\
  \times v(G_x,y-y',z-z')\mathrm{e}^{iG_x x'}\mathrm{d}y'\mathrm{d}z'.
\end{multline}
Invoking the charge neutrality along the chain axis, we have  that the
difference between electron and ionic densities satisfies
\begin{equation}
  \iint\left[n_{\rm ion}(G_x=0,y,z)-n_{\rm el}(G_x=0,y,z)\right]
  \mathrm{d}y\mathrm{d}z = 0.
\end{equation}
Unfortunately, the cutoff function in Eq.~(\ref{1DCutoff}) is divergent for
$G_x=0$. So the effective potential results in an undetermined $0\cdot\infty$
form. However, we can work out an analytical expression for it by defining first
a finite cylindrical cutoff, but then bringing the size of the cylinder to
infinity. This way, as a first step, we get a new cutoff interaction in a {\em
finite} cylinder of radius $R$, and length $h$, assuming that $h$ is much larger
than the cell size in the periodic direction. In this case the modified finite
cutoff potential includes a term
\begin{equation}
  \tilde v^{1D}(G_x,r,h)\propto\log\left(\frac{h+\sqrt{h^2+r^2}}{r}\right),
\end{equation}
which, in turn gives, for the particular plane $G_x=0$,
\begin{multline}
\tilde v^{1D}(G_x=0,G_{\bot})\approx -
4\pi\int_0^R r J_0(G_{\bot}r)\log(r)\mathrm{d}r \\ + 4\pi R\log(2h)
\frac{J_1(G_{\bot}R)}{G_{\bot}}.\label{cyl-finite} 
\end{multline}
The effective potential is now split into two terms, but only the second one
depends on $h$. The second step is achieved by going to the limit $h\to+\infty$,
to obtain the exact infinite cutoff. By calculating this limit, we notice that
only the second term in the right hand side of Eq.~(\ref{cyl-finite}) diverges.
This term is the one that can be dropped due to charge neutrality (in fact it
has the same form for the ionic and electronic charge densities). Thus, for the
cancellation to be effective in a practical implementation, we have to treat on
the same way both the ionic and Hartree Coulomb contributions. Of course the
first term in the right hand side Eq.~(\ref{cyl-finite}) has always to be taken
into account, affecting both the long and the short range part of the cutoff
potentials. 

Following this procedure, we are able to get a considerable computational
advantage, compared, e.g., to the method originally proposed by Spataru {\em et
al.},\cite{Spataru04b} since our cutoff is just an analytical function of the
reciprocal space coordinates, and the evaluation of an integral for every value
of $G_x,G_{\perp}$ is not needed. The cutoff proposed in
Ref.~\onlinecite{Spataru04b} is actually a particular case of our cutoff,
obtained by using the finite cylinder for all the components of the ${\mathbf
G}$ vectors: in this case the quadrature in Eq.~(\ref{cyl-finite}) has to be
evaluated for each $G_x$, $G_y$, and $G_z$, and a convergence study in $h$ is
mandatory (see discussion in Sec.~\ref{static-pol-section}, and Fig.
\ref{fig-pol}).

In the 1D-periodic case, the $G=0$ value is now well defined, and it turns out to be
$\lim_{G_{\bot}\to0}\tilde v(G_x,G_{\bot})$
\begin{equation}\label{cyl-finite-zero}
  \tilde v^{1D}(G_x=0,G_{\bot}=0) = - \pi R^2(2\log(R)-1).
\end{equation}

The analogous result for the 2D-periodic cutoff is obtained by imposing finite
cutoff sizes $h_x=\alpha h_y=h$ (much larger than the cell size), in the
periodic directions $x$ and $y$, and dropping the $h$-dependent part before
passing to the limit $h\to +\infty$. The constant $\alpha$ is the ratio
$\frac{L_x}{L_y}$ between the in-plane lattice vectors.
\begin{multline}
  \tilde v^{2D}(G_{\|}=0,G_z)\approx
  \frac{4\pi}{G^2_z}
    \left[
      1 - \cos(G_z R) - G_z R\sin(G_z R)
    \right]\label{pla-finite} \\
  + 8h\log\left(\frac{(\alpha+\sqrt{1+\alpha^2})(1+\sqrt{1+\alpha^2})}{\alpha}\right)\frac{\sin(G_z R)}{G_z}.
\end{multline}
The $G=0$ value is
\begin{equation}
  \tilde v^{2D}(G_{\|}=0,G_z=0) = - 2\pi R^2
\end{equation}

To summarise, the divergences can be cancelled also in 1D-periodic and
2D-periodic systems provided that 1) we apply the cutoff function to both the
ionic and the electronic potentials, 2) we separate out the infinite contribution as shown above, and 3) we properly account for the short range contributions as stated
in Table \ref{TabCutoff}.
The analytical results of the present work
are condensed in Tab. \ref{TabCutoff}: all the possible values for 
the cutoff functions are listed there as a quick reference for
the reader.


\begin{table}
  \centering
  \begin{tabular}{|c|c||l|}
    \hline
    $G$   & & 0D-periodic $\tilde v^{0D}(G) =$                     \\ 
    \hline
    $>0$  & & $\phantom{\bigg[}\frac{4\pi}{G^2}[1-\cos(GR)]$  \\ 
    \hline
    $0$   & & ${\phantom{\bigg[}2\pi R^2}$                      \\ 
    \hline  \hline
    $G_x$ & $G_{\bot}$ &  1D-periodic $\tilde v^{1D}(G_x,G_{\bot})=$           \\ 
    \hline
    $>0$  & any        & 
    \begin{tabular}{r}
    $\phantom{\bigg[}\frac{4\pi}{G^2}\big[1+G_{\bot}RJ_1(G_{\bot}R)K_0(G_x R)$  \\
    $-G_x RJ_0(G_{\bot}R)K_1(G_x R)\big]\phantom{\bigg]}$
    \end{tabular}                       \\ 
    \hline
    $0$   & $>0$       & ${\phantom{\bigg[}-4\pi\int_0^R r J_0(G_{\bot}r)\log(r)\mathrm{d}r}$                            \\ 
    \hline
    $0$   & $0$        & ${\phantom{\bigg[}-\pi R^2(2\log(R)-1)}$  \\ 
    \hline \hline
    $G_{\|}$ &  $G_z$ & 2D-periodic  $\tilde v^{2D}(G_{\|},G_z) =$     \\ 
    \hline
    $>0$     & any   &   $\phantom{\bigg[}\frac{4\pi}{G^2}\left[1+\mathrm{e}^{-G_{\|} R}\left(\frac{G_z}{G_{\|}}\sin(G_z R)-\cos(G_z R)\right)\right]$ \\ 
    \hline
    $0$      & $>0$        &     ${\phantom{\bigg[}\frac{4\pi}{G^2_z}\left[1 - \cos(G_z R) - G_z R\sin(G_z R)\right]}$   \\ 
    \hline
    $0$      & $0$         &     ${\phantom{\bigg[}- 2\pi R^2}$   \\ 
    \hline
  \end{tabular}
  \caption{Reference Table summarising the results of the cutoff work for charge-neutral systems: finite systems (0D), one-dimensional systems (1D) and two-dimensional
  systems (2D). The complete reciprocal space expression of the Hartree potential
  is provided. For the 1D case, $R$ stands for the radius of the cylindrical cutoff 
  whereas in the 0D case is the radius of the spherical cutoff. In 2D stands
  for half the thickness of the slab cutoff (see text for details).
  \label{TabCutoff}}
\end{table}

\section{Results}\label{sect-results}

The scheme illustrated above has been implemented both in the real space
time-dependent DFT code OCTOPUS\cite{octopus}, and in the plane wave 
many-body-perturbation-theory (MBPT) code SELF\cite{self}. The tests have been
performed on the prototypical cases of infinite chains of atoms along the $x$
axis. The comparisons are performed between the 3D-periodic calculation
(physically corresponding to a crystal of chains), and the 1D-periodic case
(corresponding to the isolated chain) both in the usual supercell approach, and
within our exact screening method. The discussion for the 2D cases follows the
same path as for the 1D case, while results for the finite systems have already
been reported in the literature.\cite{Jarvis97, Castro03} We addressed different
properties to see the impact of the cutoff at  each level of calculation, from
the ground state to excited state and  quasiparticle dynamics.

\subsection{Ground state calculations}

All the calculations have been done with the real-space implementation of DFT in
the OCTOPUS\cite{octopus} code. We have used non-local norm-conserving
pseudopotentials\cite{trouiller} to describe the electron-ion interaction  and
the local-density approximation (LDA)\cite{pz} to describe exchange-correlation
effects. The particular choice of exchange-correlation or ionic-pseudopotential
does not matter here as we want to assert the impact of the Coulomb cutoff and
this is independent of those quantities. Moreover, we have used a grid of 0.38
a.u. for Si and Na.

In this case the footprint of the interaction of neighbouring chains in the $y$
and $z$ direction is the dispersion of the bands in the corresponding direction
of the Brillouin zone. However it is known that, if the supercell is large
enough, the bands along the $\Gamma-X$ direction are unchanged. This is in
apparent contradiction with the fact that the radial ionic potential for a wire
(that asymptotically goes like $\ln(r)$ as a function of the distance $r$ from
the axis of the wire) is completely different from the crystal potential.

The answer to this contradiction is clear if we perform a cutoff calculation. In
fact the overall effect on the occupied states turns out to be cancelled by the
Hartree potential, i.e. by the electron screening of the ionic potential, but
two different scenarios are visible as soon as the proper cutoff is used.

\begin{figure}
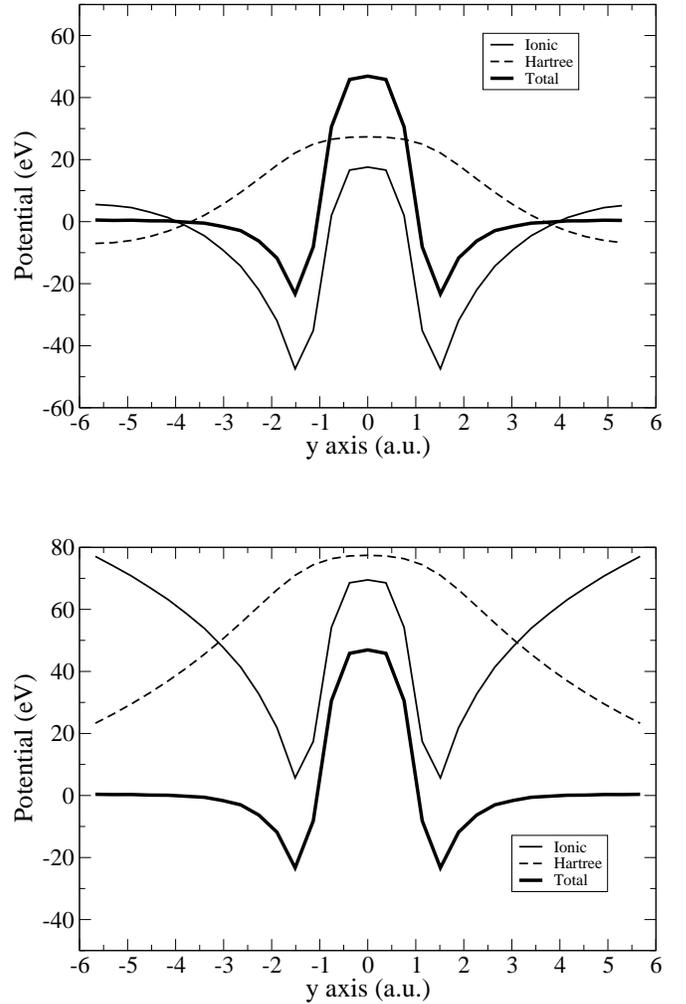

\begin{tabular}{c} 
\epsfig{file=fig/Si.3.pot.y.eps,width=\columnwidth} \\ \\ \\
\epsfig{file=fig/Si.1.pot.y.eps,width=\columnwidth}
\end{tabular} \caption{Calculated total and ionic and Hartree potentials
for a 3D-periodic (top) and 1D-periodic (bottom) Si chain}.\label{fig-SiPot}
\end{figure}

In Fig. \ref{fig-SiPot} (top) it is shown the ionic potential, the Hartree
potential, and their sum for a Si atom in a parallelepiped supercell with side
lengths of 2.5, 11, and 11 a.u. respectively in the $x$, $y$ and $z$ directions.
No cutoff is used here. The ionic potential is roughly behaving like
$\frac{1}{r}$ in the area not too close to the nucleus (where the pseudopotential takes over). The total potential, on the other hand, falls off
rapidly to an almost constant value at around 4 a.u. from the nuclear position,
by effect of the electron screening.

Fig. \ref{fig-SiPot} (bottom) shows the results when the cutoff is applied
(the radius of the cylinder is $R=5.5$~a.u. such that there is zero interaction
between cells). The ionic potential now behaves like it is expected for a
potential of a chain, i.e. diverges logarithmically, and is clearly different
from the latter case. Nevertheless the sum of the ionic and Hartree potential is
basically the same as for the 3D-periodic system.

\begin{figure}
    \centering
    \epsfig{file=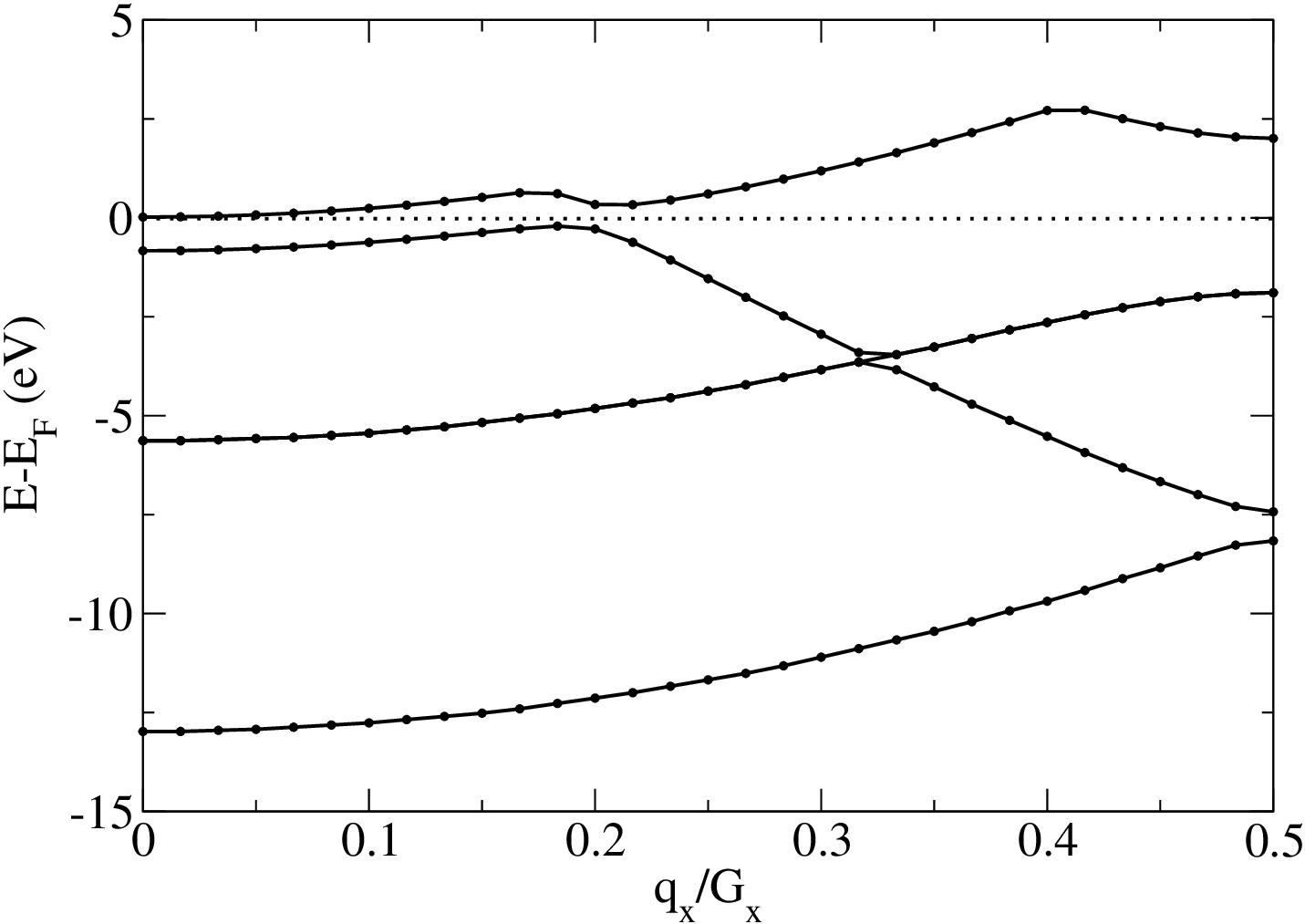,width=\columnwidth}
    \caption{Si linear chain in a supercell size of 4.9x19x19 a.u.\label{fig-SiBands}}
\end{figure}

In the static case the two band structure are then expected, and are found to be
the same, confirming that, as far as static calculations are performed, the
supercell approximation is good, provided that the supercell is large enough
(see Fig.~\ref{fig-SiBands}). In static calculations, then, the use of our
cutoff only has the effect of allowing us to eventually use a smaller supercell,
what provides clear computational savings. In the case of the Si-chain a full 3D
calculation would need of a cell size of 38 a.u. whereas the cutoff calculation
would give the same result with a cell size of 19~a.u. Of course, when more delocalized states are considered, like higher energy unoccupied states, larger differences are observed with respect to the supercell calculation.

\begin{figure}
    \centering
    \epsfig{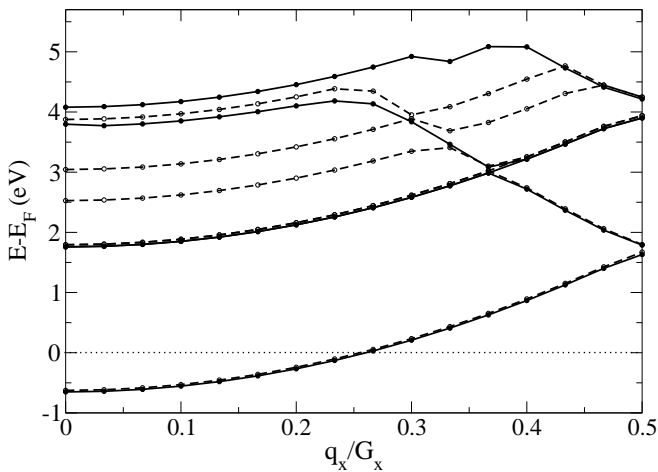}
    \caption{Effect of the cutoff in a Na linear chain
    in a supercell size of 7.5x19x19 a.u. The bands obtained with an ordinary supercell calculation with no cutoff (dashed line) are compared to the bands obtained applying the 1D cylindrical cutoff (solid line). As it is explained in the text, only the unoccupied levels are affected by the cutoff.\label{fig-NaBands}}
\end{figure}

In Fig.~\ref{fig-NaBands} a Na chain with lattice constant 7.5 a.u. is
considered in a cell of 7.5x19x19 a.u., and the effect of the cutoff on the
occupied and unoccupied stated is shown. As expected, the occupied states are
not affected by the use of the cutoff, since the density of the system within
the cutoff radius is unchanged, and the corresponding band is the same as it is
found for an ordinary 3D supercell calculation with the same cell size. However
there is a clear effect on the bands corresponding to unoccupied states, and the
effect is larger the higher is the energy of the states. In fact the high energy
states, and the states in the continuum are more delocalized, and therefore the
effect of the boundary conditions is more sensible.

\subsection{Static polarisability}\label{static-pol-section}

After the successful analysis of the ground state properties with the cutoff
scheme, we have applied the modified Coulomb potential to calculate the static
polarisability of an infinite chain in the Random Phase Approximation (RPA). As
a test case we have considered a chain made of hydrogen atoms, two atoms per
cell at a distance of 2 a.u. The lattice parameter was 4.5 a.u. For this system
we have also calculated excited state properties in many-body perturbation
theory, in particular the quasiparticle gap in Hedin's GW approximation
\cite{GW} and the  optical absorption spectra in the Bethe-Salpeter
framework\cite{Onida02,bse} (see subsections below). All these calculations have
been performed in the code SELF.\cite{self} The polarisability for the monomer,
{\em i.e. a finite system}, in the RPA approximation  including local field
effects is defined as
\begin{equation}\label{alpha}
  \alpha=-\lim_{{\bf q \rightarrow 0}} \frac{1}{q^2}\chi^{\phantom 0}_{00}({\bf q})\frac{\Omega}{4\pi}.
\end{equation}
where $\chi^{\phantom 0}_{GG'}(q)$ is the interacting polarisation function that is solution of the Dyson like equation
\begin{equation}\label{dyson}
  \chi^{\phantom 0}_{ GG'}({\bf q})=  \chi^{0}_{GG'}({\bf q}) 
      +\sum_{G''}\chi^{0}_{GG''}({\bf q}) v({\bf q+G''}) \chi^{\phantom 0}_{ G''G'}({\bf q}).
\end{equation}
and $\chi^{0}$ is the non interacting polarisation function obtained by the
Adler-Wiser expression.\cite{adler} v({\bf q}+{\bf G}) are the Fourier
components of the Coulomb interaction. Note that the expression of  $\alpha$  in
Eq.~\ref{alpha} is also valid for calculations in finite systems, in the
supercell approximation, and the dependence from the wave-vector $q$ is due to
the representation in reciprocal space.

\begin{figure}
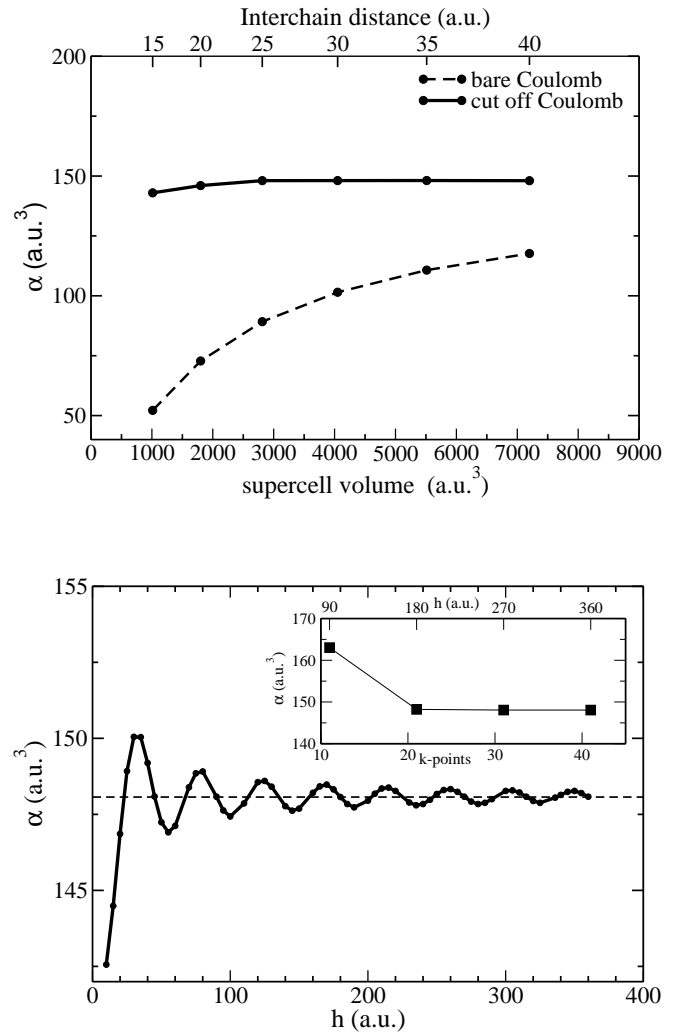

\begin{tabular}{c} 
  \epsfig{file=fig/alpha-convergence.eps,width=\columnwidth} \\ \\ \\
  \epsfig{file=fig/finite-vs-exact-1D-cutoff.eps,width=\columnwidth} 
\end{tabular}
\caption{Top: Polarisability per unit cell of an $H_2$ chain in RPA
approximation as a function of the supercell volume. The solid line joins the
values obtained with the cutoff potential, while the dashed lines joins the
values obtained with the bare Coulomb potential. The cutoff radius is 8.0 a.u.
The inter-chain distance is indicated in the top axis. Bottom: Polarisability of
the $H_2$ chain calculated with the finite cutoff potential of
Ref.\onlinecite{Spataru04b}. In abscissa different values of the cutoff length
along the chain axis. The dashed straight line indicates the value obtained with
the cutoff of Eq.(\ref{1DCutoff}).In the inset we show the convergence of the
polarisability with respect to the k-points sampling along the chain axis obtained with the cutoff of Eq.(\ref{1DCutoff}). In the upper axis it is indicated the maximum allowed length $h$ for each k-point sampling used in the calculation of the $G_x=0$ components by Eq.(\ref{cyl-finite}).\label{fig-pol}}
\end{figure}

In the top panel of Fig.~\ref{fig-pol} we compare the values of the calculated
polarisability $\alpha$ for different supercell sizes. $\alpha$ is calculated
both using the bare Coulomb $v({\bf q+G})=\frac{4\pi}{|{\bf q+G}|^2}$ and the
modified cutoff potential of Eq.(\ref{1DCutoff}) (the radius of the cutoff is
always set to half the inter-chain distance). The lattice constant along the
chain axis is kept fixed. Using the cutoff the static polarisability already
converges to the asymptotic value with an inter-chain distance of 25 a.u., while
without the cutoff the convergence is much slower, and the exact value is
approximated to the same accuracy for much larger cell sizes (beyond the 
calculations shown in the top of Fig.~\ref{fig-pol}). 

We must stress that the treatment of the divergences in this case is different
with respect to the case of the Hartree and ionic potential cancellation
for ground-state calculations (i.e. charge neutrality). In fact, while in
the calculation for the Hartree and ionic potential the diverging terms are
simply dropped by virtue of the neutralising positive background, here the
$h$-dependence in Eq.~(\ref{cyl-finite}) can be removed only for the head
component by virtue of the vanishing limit $lim_{{\bf q} \to
0}\chi^{0}_{00}({\bf q})=0$, while  for the other $G_x=0$ components we have to
resort to the expression of the finite cylindrical cutoff as in
Eq.(\ref{cyl-finite}).

A finite version of the 1D cutoff has been recently applied to nanotube
calculations.\cite{Spataru04,Spataru04b} This cutoff was obtained by numerically
truncating the Coulomb interaction along the axis of the nanotube, in addition
to the radial truncation. Therefore the effective interaction is limited to a
finite cylinder, whose size can be up to a hundred times the unit cell size,
depending on the density of the k-point sampling along the axis.\cite{maxh} The cutoff axial length has to be larger than the expected bound exciton length.

In the bottom part of Fig.\ref{fig-pol} we compare the results obtained with our
analytical cutoff (Eq.~(\ref{1DCutoff})) with its finite special case as
proposed in Ref.~\onlinecite{Spataru04b}. We observe that the value of the
static polarisability calculated with the finite cutoff oscillates around an
asymptotic value, for increasing axial cutoff lengths. The asymptotic value
exactly coincides with the value that is obtained with our cutoff. We stress
that we also resort to the finite form of the cutoff only for the diverging of
components of the potential, thus we note that there is a clear numerical
advantage in using our expression, since the cutoff is analytical for all values
except at $G_x=0$, and the corresponding quadrature has to be numerically
evaluated for these points only. In the inset of the bottom part of
Fig.~\ref{fig-pol} it is also shown the convergences of the polarisability
obtained with our cutoff with respect to the k-points sampling. The sampling is
unidimensional along the axial direction. Observe that the calculation using our
cutoff is already converged for a sampling of 20 k-points. In the upper axis it
is also indicated the corresponding maximum allowed value of  the finite cutoff
length in the axial direction that has been used to calculate the $G_x=0$
components.

Finite size effects turn out to be relevant also for many-body perturbation
theory calculations. For the same test system (linear H$_2$-chain), in the next
two subsections, we consider the performance of our cutoff potential for the
calculation of the quasiparticle energies in GW\cite{GW} approximation and in
the absorption spectra  in the Bethe-Salpeter framework.\cite{Onida02,bse}

\subsubsection{Quasiparticles in the GW approximation}

In the GW approximation, the non-local energy-dependent electronic self-energy
$\Sigma$ plays a role similar to that of the exchange-correlation potential of
DFT. $\Sigma$ is approximated by the  convolution of the one electron Green's
function and the dynamically screened Coulomb interaction $W$. We first
calculate the ground state electronic properties using the DFT code
ABINIT.\cite{abinit} These calculation are performed in LDA\cite{pz}, and 
pseudopotentials\cite{trouiller} approximation. An energy  cutoff of 30
hartree has been used to get converged results.  The LDA eigenvalues and 
eigenfunctions are then used to construct the RPA screened Coulomb interaction
$W$, and the GW self-energy. The inverse dielectric matrix
$\epsilon^{-1}_{G,G'}$ has been calculated  using the plasmon-pole approximation
\cite{plasmonpole} and the quasiparticle energies have been calculated at the
first order of perturbation theory in $\Sigma-V_{xc}$.\cite{Hybertsen} Dividing
the self-energy in an exchange $\Sigma_x$ and a correlation  $\Sigma_c$ parts
($\langle \phi_j^{DFT}|\Sigma|\phi_i^{DFT}\rangle=   \langle
\phi_j^{DFT}|\Sigma_x|\phi_i^{DFT}\rangle+ \langle
\phi_j^{DFT}|\Sigma_c|\phi_i^{DFT}\rangle$), we get the following representation
for the self-energy in a plane-waves basis set:
\begin{multline}
\langle n {\bf k}|\Sigma_x({\bf r_1,r_2})| n' {\bf k'}\rangle
 = -\sum_{n_1}\int_{Bz}\frac{d^3{\bf q}}{(2\pi)^3} \sum_G v(\bf{q+G}) \times \\ 
 \times \rho^{\phantom \star}_{nn_1}({\bf q,G})\rho_{n'n_1}^{\star}({\bf q,G})f_{n_1{\bf k_1}}
\end{multline}
and
\begin{multline}
\langle n {\bf k}|\Sigma_c({\bf r_1,r_2},\omega)| n' {\bf k'}\rangle = 
\frac{1}{2}\sum_{n_1} \int_{Bz}\frac{d^3{\bf q}}{(2\pi)^3}{\Bigg\{}\sum_{{\bf GG'}} v(\bf{q+G'}) \label{selfenergy}\\
\times \rho^{\phantom \star}_{nn_1}({\bf q,G})\rho_{n'n_1}^{\star}({\bf q,G'})  
\int \frac{d\omega'}{2\pi} \epsilon^{-1}_{{\bf GG'}}({\bf q},\omega') \\
\times {\Big[}\frac{f_{n1({\bf k-q})}}{\omega-\omega'-\epsilon^{LDA}_{n1({\bf k-q})}-i\delta}
+ \frac{1-f_{n1({\bf k-q})}}{\omega-\omega'-\epsilon^{LDA}_{n1({\bf k-q})}+i\delta}\Big{]}
{\Bigg\}}
\end{multline}
where $\rho_{nn_1}({\bf q+G})= \langle n {\bf k}|e^{i({\bf q+G}) \cdot {\bf
r_1}}| n_1 {\bf k_1}\rangle $ and  the integral in the frequency domain in
Eq.(\ref{selfenergy}) has been analytically solved considering the dielectric
matrix in the plasmon pole mode: ( $\epsilon^{-1}_{{\bf
G,G'}}(\omega)=\delta^{\phantom{-1}}_{\bf{G,G'}}+\Omega^{\phantom{-1}}_{{\bf
G,G'}}/(\omega^2-\tilde{\omega}^{2}_{G,G'})$).

In order to eliminate the spurious interaction between different supercells,
leaving the bare Coulomb interaction unchanged along the chain direction, we
just introduce the expression of Eq.(\ref{1DCutoff}) in the construction of
$\Sigma_x$ and $\Sigma_c$, and also in the calculation of $\epsilon^{-1}_{GG'}$.
As we did for the calculation of the static polarisability, 
the divergences appearing in the  components $(G_x=0)$ cannot be fully
removed and for such components we resort to the finite version of
the cutoff potential Eq.(\ref{cyl-finite}).

\begin{figure}[t]
    \centering
    \epsfig{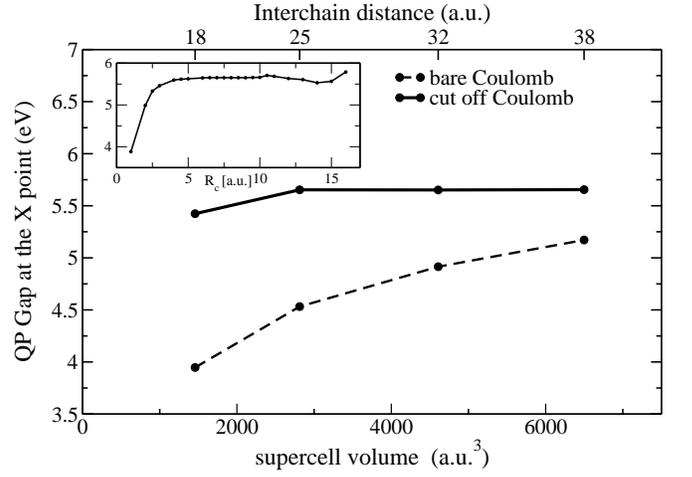}
    \caption {Convergence of the GW quasiparticle gap for the $H_2$ chain as a function of the cell size, using the bare Coulomb potential (dashed line) and the cutoff potential (solid line).In the inset the behaviour of the GW quasiparticle gap as a function of the value of the cutoff radius for a supercell with inter-chain distance of 32 a.u. is shown. The plateau obtained around a radius of 8 a.u. (i.e. one fourth of the supercell size) corresponds to the situation in which the radial images of chains no longer mutually interact, and the calculation is converged. Increasing the radius above approximately 12 a.u. the interaction is back and produces oscillations in the value of the gap.
\label{fig-GWgap}}
\end{figure}

In Fig. \ref{fig-GWgap} the convergence of the quasiparticle gap at the $X$
point is calculated for different supercell sizes in the GW approximation. A
cutoff radius of 8.0 a.u. has been used. When the cutoff potential is used, 60 k
points in the axis direction has been necessary to get converged results. In the
inset of Fig. \ref{fig-GWgap} we show the behaviour of the quasiparticle gap in
function of the cutoff radius. We observe that for $R_c > 6$~a.u. a plateau is
reached, and, for $R_c > 12$~a.u., a small oscillation appears due to
interaction between the tails of the charge density of the system with its image
in the neighbour cell. Differently from the DFT-LDA, calculation for neutral
systems, where the supercell approximation turns out to be good, as we have
discussed above, we can see that the convergence of the GW quasiparticle
correction turns out to be extremely slow with respect to the size of the
supercell and huge supercells are needed in order to get converged results. This
is due to the fact that in the GW calculation the addition of an electron (or a
hole) to the system induces charge oscillation in the periodic images too. It is
important to note that the slow convergence is caused by the correlation part of
the self-energy (Eq.(\ref{selfenergy})), while the exchange part is rapidly
convergent with respect to the cell size. The use of the cutoff Coulomb
potential really improves drastically the convergence as it is evident from
Fig.\ref{fig-GWgap}. Notice that still at 38~a.u. inter-chain distance the GW
gap is underestimated by about 0.5~eV.  A similar trend (but with smaller
variations) has been found by Onida {\it et al.} \cite{Onida95}, for a finite
system (Sodium Tetramer) using the cutoff potential of Eq.(\ref{0Dcutoff}). 
Clearly there is a strong dimensionality dependence of the self-energy
correction.  The non-monotonic behaviour versus dimensionality of the
self-energy  correction has also been pointed out in Ref.~\onlinecite{delerue}
where the gap-correction was shown to have a strong component of the surface
polarisation.

\subsubsection{Exciton binding energy: Bethe-Salpeter equation}

Starting from the quasiparticle energies we have calculated the optical 
absorption spectra including electron-hole interactions solving the
Bethe-Salpeter equation \cite{bse}. The basis set to describe the exciton state
is composed  by product states of the occupied and unoccupied LDA single
particle states and the coupled electron-hole excited states $|S\rangle = \sum_{cv{\bf
k}} A_{cv{\bf k}} a^\dag_{c{\bf k}}a_{v{\bf k}}|0\rangle$, where  $|0\rangle$ is the ground
state of the system. $A_{cv{\bf k}}$ is the probability amplitude of finding 
an excited electron in the state $(c{\bf k})$ and a hole in $(v{\bf k})$, and
it satisfies the equation
\begin{equation}\label{eq-bse}
(\epsilon^{QP}_{c{\bf k}}-\epsilon^{QP}_{v{\bf k}})A_{vc{\bf k}}+\sum_{vc{\bf k}, v'c'{\bf k'}}
K_{vc{\bf k}, v'c'{\bf k'}}A_{v'c'{\bf k'}}=E_S A_{vc{\bf k}}
\end{equation}
$E_S$ is the excitation energy of the state $|S\rangle$ and $K$ the interaction
kernel that includes an unscreened exchange repulsive term $K^{Exch}$  and a
screened  electron-hole interaction $K^{dir}$ (direct term).   In plane
wave basis such terms read
\begin{equation}
K^{Exch}_{{vc{\bf k} \choose v'c'{\bf k'}}} = \frac{2}{\Omega}\sum_{{\bf G \neq 0}} v({\bf G})
\langle c{\bf k} |e^{i{\bf G\cdot r}}|v {\bf k} \rangle 
  \langle v'{\bf k'} |e^{-i{\bf G'\cdot r}}|c' {\bf k'} \rangle 
\label{exchange}
  \end{equation}
\begin{align}
K^{dir}_{{vc{\bf k} \choose v'c'{\bf k'}}} = & \frac{1}{\Omega}\sum_{{\bf G,G'}} v({\bf q+G})
\epsilon^{-1}_{{\bf GG'}}({\bf q}) \langle c{\bf k} |e^{i{\bf (q+G)\cdot r}}|c' {\bf k'} \rangle \times \nonumber \\
& \times \langle v'{\bf k'} |e^{-i{\bf (q+G')\cdot r}}|v {\bf k} \rangle \delta_{\bf{q.k-k'}}\nonumber  \\
\label{direct}
\end{align}
The screened potential has been treated in static RPA approximation  (dynamical
effects in the screening have been neglected as it is usually done in present
Bethe-Salpeter calculations\cite{bse}). The quasiparticle energies entering in
the diagonal part of the  Hamiltonian in Eq.~(\ref{eq-bse}) are obtained
applying a scissor  operator to the LDA energies, because in the studied test
case the main difference between the quasiparticle and LDA band structure
consists of a rigid energy shift of energy bands. From the solution of the
Bethe-Salpeter equation (Eq.\ref{eq-bse}) it is possible to calculate the 
macroscopic dielectric function, in particular the imaginary part reads
\begin{equation}
\epsilon_2(\omega)=\frac{1}{\Omega}\frac{4\pi^2e^2}{\omega^2}\sum_S\Bigg| \sum_{vc{\bf k}}A^S_{vc{\bf k}}
\langle v{\bf k}|\boldsymbol{\lambda} \cdot \boldsymbol{\nu} | c{\bf k} \rangle \Bigg |^2 \delta(E_S-\hbar\omega)
\end{equation}
where the summation runs over all the vertical excitations from the ground state
$|0\rangle$ to the excited state $|S\rangle$, $E_S$ is the corresponding
excitation energy, $\boldsymbol{\nu}$ is the  velocity operator and
$\boldsymbol{\lambda}$ is the polarisation vector. As in the case of GW
calculation, in order to isolate the chain, we substitute the cutoff potential
of Eq.(\ref{1DCutoff}) both in the exchange term Eq.(\ref{exchange}) and in the
direct term of the Bethe-Salpeter equation Eq.(\ref{direct}), as well as for the
RPA dielectric matrix present in Eq.(\ref{direct}).

\begin{figure}
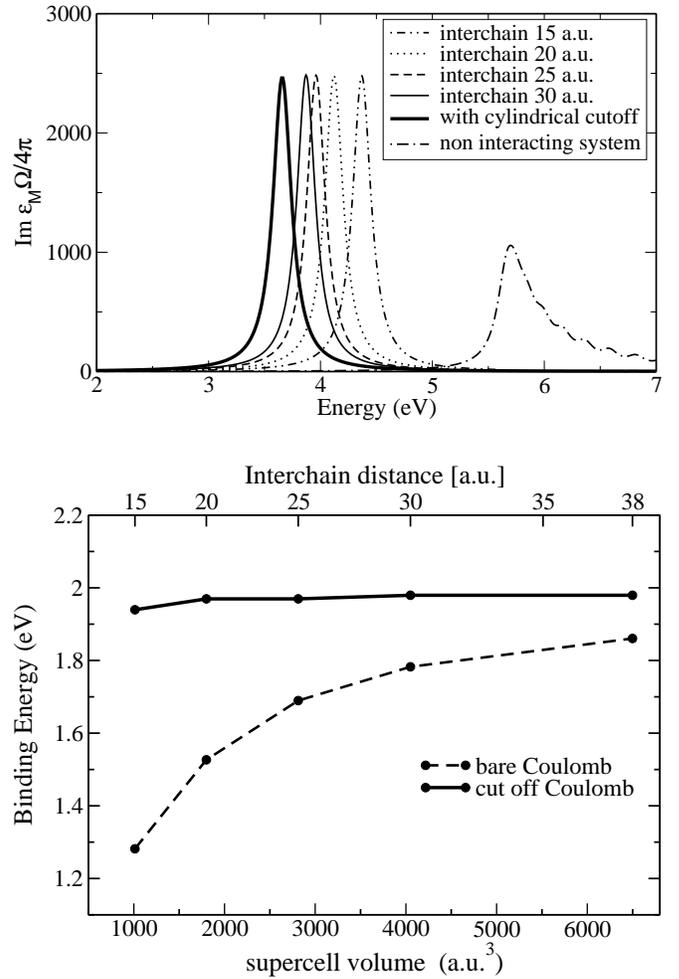

\centering
\begin{tabular}{c}
  \epsfig{file=fig/BSE-photoabs-convergence.eps,width=\columnwidth} \\
  \\
  \epsfig{file=fig/BSE-binding-energy-convergence.eps,width=\columnwidth}
\end{tabular}
\caption{Top: Photo absorption cross section for different supercell volumes. In the legend the inter-chain distances corresponding to each volume are indicated. The intensity have been normalised to the volume of the supercell. The non-interacting absorption spectra and the spectra obtained with the cutoff potential are also included. Bottom: exciton binding energy vs supercell volume calculated using the cut off potential (solid line), and the bare Coulomb potential (dashed line).} \label{fig-bse}
\end{figure}

In the top part of Fig. \ref{fig-bse} we show the calculated spectra for
different cell sizes together with the non-interacting spectrum, and the
spectrum obtained using the cutoff Coulomb potential for an inter-chain distance
larger than 20 a.u. and cylindrical cutoff radius of 8 a.u. The scissor operator
applied in this calculation is the same for all the volumes and correspond to
the converged GW gap. 

As it is known, the electron-hole interaction modifies both the shape and the
energy of the main absorption peak. This effect is related to the the slow
evolution of the polarisability per $H_2$ unit\cite{daniele}. Furthermore, the
present results clearly illustrate that the spectrum calculated without the
cutoff slowly converges towards the exact result. This is highlighted in the
bottom panel of Fig. \ref{fig-bse} where we show the dependency of the exciton
binding energy on the supercell volume, the binding energy being defined as the
energy difference between the excitonic peak and the optical gap. We observe
that the effect of the inter-chain interaction consists in reducing the binding
energy with respect to its value in the isolated system. This value is slowly
approached as the inter-chain distance increases, while, once the cutoff is
applied to the Coulomb potential, the limit is reached as soon as the densities
of the system and its periodical images do not interact. If we consider the
convergence of the quasiparticle gap and of the binding energy with respect to
the cell volume we notice that, if a cutoff is not used, the position of the
absorption peak is controlled by the convergence of the Bethe-Salpeter equation
solution, which, in turn, depends on the (slower) convergence of the GW
energies. It is clear from Fig.~\ref{fig-GWgap} that the use of the cutoff
allows us to considerably speed up this bottleneck.

\begin{figure}
    \centering
\begin{tabular}{c}
    \epsfig{file=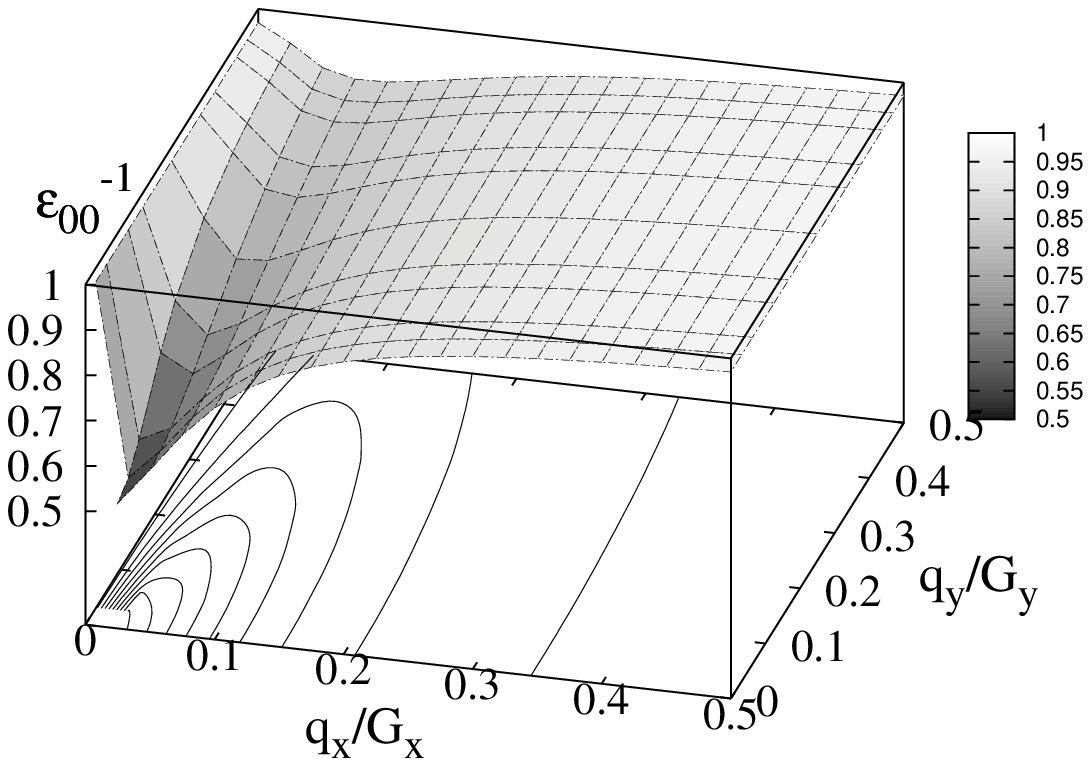,width=\columnwidth}  \\
    \epsfig{file=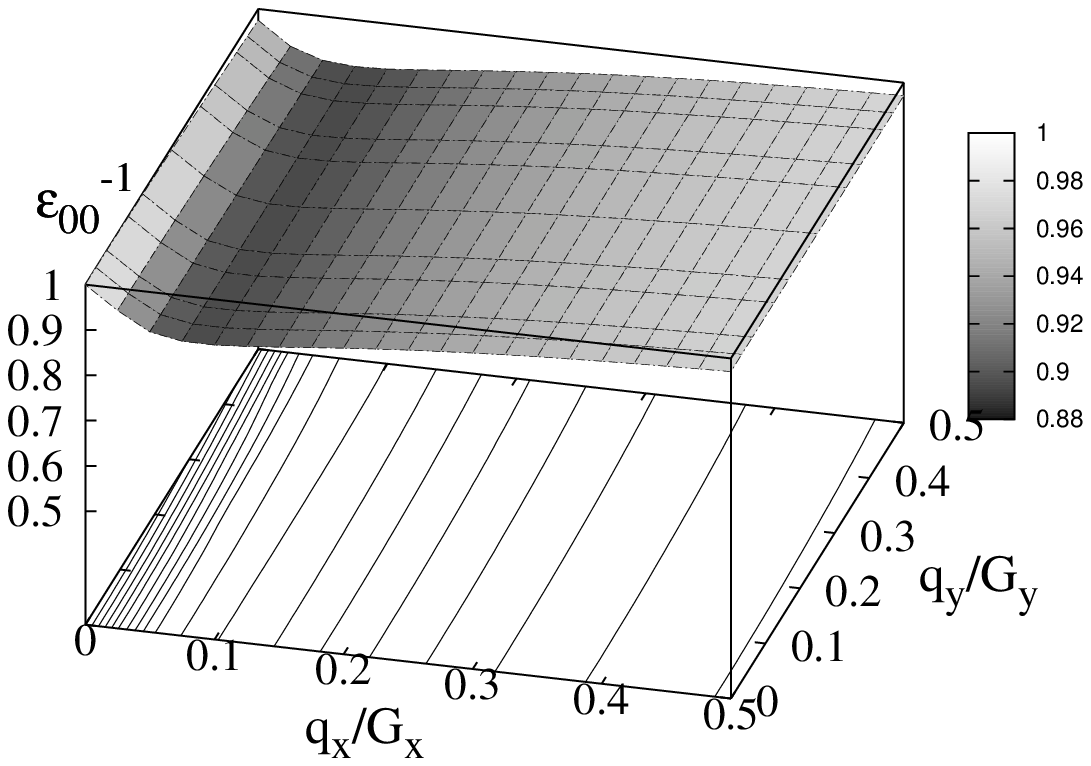,width=\columnwidth} 
\end{tabular}
     \caption{Values of $\epsilon^{-1}_{00}(q_x,q_y)$ in the $q_z=0$ plane for the $H_2$ chain with an inter-chain distance of 25 a.u., using the bare Coulomb (top) and the cutoff potential (bottom). The axis of the chain is along the x direction} 
\label{fig-epsilon}
\end{figure}

The use of our cutoff also has an important effect on the Brillouin zone
sampling. In Fig. \ref{fig-epsilon} we show the value of
$\epsilon^{-1}_{00}(q_x,q_y)$ in the $q_z=0$ plane for a supercell corresponding
to an inter-chain distance of 25 a.u. When the cutoff potential is used (bottom
panel) the screening is smaller, compared to the case of the bare potential (top
panel). Looking at the direction perpendicular to the chain (the chain axis is
along the $x$ direction) we see that the dielectric matrix is approximately
constant, and this fact allows us to sample the Brillouin zone only in the
direction of the chain axis. For both the GW and Bethe-Salpeter calculations a
three dimensional sampling of the Brillouin zone is needed to get converged
results when no cutoff is used, while a simple one-dimensional sampling can be
adopted when the interaction is cutoff.

\section{Conclusions}\label{sect-concl}

An infinite system is an artifact that allows us to exploit the powerful
symmetry properties of an ideal system to approximate the properties of a finite
one that is too large to be simulated at once. In order to use the valuable
supercell approximation for systems that are periodic in less than three
dimensions some cutoff technique is required. The technique presented here
provides a recipe to build the supercell, and a truncated Coulomb potential in
Fourier space in such a way that the interactions to all the undesired images of
the system are cancelled. This technique is exact for the given supercell sizes,
and can be in most cases well approximated using supercells whose length is the
double of the system size in the non-periodic dimensions. The method is very
easily implemented in all available codes that use the supercell scheme, and is
independent on the adopted basis set. We have tested it both in a real space
code for LDA band-structure calculation of an atomic chain, and a plane wave
code, for the static polarisability in RPA approximation, GW quasiparticle
correction, and photo-absorption spectra in the Bethe-Salpeter scheme, showing
that the convergence with respect the vacuum needed to isolate the system from
its images is greatly enhanced, and the sampling of the Brillouin zone is
heavily reduced, being only necessary along the periodic directions of the
system.


\begin{acknowledgments} 
This research was supported by EU Research and Training Network ``Exciting"
(contract HPRN-CT-2002-00317),  EC 6th framework Network of Excellence
NANOQUANTA (NMP4-CT-2004-500198) and Spanish MEC. AR acknowledges the Foundation under the Bessel research award (2005). We thank Dr. Alberto Castro, Dr. Miguel A.L Marques, and Heiko Appel for helpful discussions and friendly collaboration.
\end{acknowledgments}



\begin{thebibliography}{99}

\bibitem{pw} M.~L. Cohen, Solid Stat. Commun. {\bf 92}, 45 (1994);
Phys. Scri. {\bf 1}, 5 (1982). J. Ihm, A. Zunger and M.~L. Cohen, J. Phys. {\bf C 12}, 4409 (1979).
W.~E. Pickett, Comput. Phys. Rep. {\bf 9}, 115 (1989); M.~C. Payne, M.~P. Teter, D.~C. Allan, T.~A. Arias and .~D. Joannopoulos, Rev. Mod. Phys. {\bf 64}, 1045 (1992).

\bibitem{fftw} M. Frigo, S.~G. Johnson,
Proc. IEEE Int. Conf. Acoust. Speech, Signal Processing, ICASSP'98, {\bf 3}, 1381 (1998)

\bibitem{Lebowitz69} J.~E. Lebowitz, E.~H. Lieb,
Phys. Rev. Lett. {\bf 22} 631 (1969)

\bibitem{nano} See for example:
A.P. Alivisatos, Science {\bf 271}, 933 (1996);
{\it Carbon Nanotubes: Synthesis, Structure,
Properties, and Applications}, M.S. Dresselhaus, G. Dresselhaus, and
Ph. Avouris (Editors), Springer Verlag (2001);
{\it Encyclopedia of Nanoscience and Nanotechnology}  and  {Handbook of Nanostructured Biomaterials and Their Applications in Nanobiotechnology}, H.~S. Nalwa (editor), American Scientific Publishers (2005);
{\it Quantum Computing and Quantum Communication}, G. Burkard, H-A. Engel, and D. Loss, http://theorie5.physik.unibas.ch/qcomp/qcomp.html

\bibitem{maxh} The maximum value of the size of the cutoff axial length $h$ 
both in  Ref.~\onlinecite{Spataru04b}  and in Eq.~(\ref{cyl-finite}) is
$2\pi/\Delta k$, where $\Delta k$ is  the spacing of the k-point grid in the
direction of the cylinder axis.

\bibitem{GW} L. Hedin, Phys. Rev. {\bf 139}, 796 (1965);
L. Hedin and S. Lundqvist, Solid State Phys {\bf 23}, 1 (1969);
F. Aryasetiawan and O. Gunnarson, Rep. Prog. Phys. {\bf 61}, 237 (1998);
W.~G. Aulbur, L. J\"onsson and J.W. Wilkins, Solid State Physics {\bf 54}, 1 (1999).

\bibitem{Onida02} G. Onida, L. Reining, A. Rubio,
Rev. Mod. Phys. {\bf 74}, 601 (2002), and references therein.

\bibitem{ijm}  Sottile, F. Bruneval, A.~G. Marinopoulos, L. Dash, S. Botti, V. Olevano, N. Vast, A. Rubio and L. Reining, Int. J. Quantum Chem. {\bf 112}, 684 (2005).


\bibitem{Wood04} B. Wood, W. M. C. Foulkes, M.~D. Towler, and N.D. Drummond, J. Phys.: Condens. Matter {\bf 16}, 891 (2004)

\bibitem{Makov95} G. Makov, M.C. Payne,
Phys Rev. {\bf B 51}, 4014 (1995)

\bibitem{DeLeeuw80} S.~W. DeLeeuw, J.~W. Perram, E.~R. Smith,
Proc. R. Soc. Lond. {\bf A 373}, 27 (1980)

\bibitem{Ewald21} P.~P. Ewald,
Ann. Phys. {\bf 64}, 253 (1921)


\bibitem{Sphor94} E. Spohr, J. Chem. Phys. {\bf 107} 6342 (1994)

\bibitem{Yeh99} I. Yeh, M. Berkowitz,
J. Chem. Phys. {\bf 111}, 3155 (1999)

\bibitem{Martyna99} G.~J. Martyna, M.~E. Tuckerman, 
J. Chem. Phys {\bf 110}, 2810 (1999)

\bibitem{Brodka03} A. Br\'odka,
Mol. Phys. {\bf B 61}, 3177 (2003)

\bibitem{Heyes77} D.~M. Heyes, M. Barber, J.~H.~R. Clarke,
J. Chem. Soc. Faraday Trans II {\bf 73}, 1485 (1977)

\bibitem{Grzybowski00} A. Grzybowski, E. Gw\'o\'zd\'z, A. Br\'odka,
Phys. Rev. {\bf B 61},  6706 (2000)

\bibitem{Minary02} P. Min\'ary, M.~E. Tuckerman, K.~A. Pihakari, and G.~J. Martyna J. Chem. Phys. {\bf 116}, 5351 (2002)

\bibitem{Minary04} P. Min\'ary, J.~A. Morrone, D.~A. Yarne, M.~E. Tuckerman, and G.~J. Martyna, J. Chem. Phys. {\bf 121}, 11949 (2004)

\bibitem{Schultz99} P.A. Schultz,
Phys. Rev. {\bf B 60}, (1999) 1551

\bibitem{Hockney} R.~W. Hockney and J.~W. Eastwood,
{\em Computer Simulations Using Particles}, Mc-Graw Hill, New York, 1981

\bibitem{Jarvis97} M.~R. Jarvis, I.~D. White, R.~W. Godby, M.~C. Payne,
Phys. Rev. {\bf B 56}, 14972 (1997)

\bibitem{fmm} L. Greengard, {\it The rapid evaluation of potential fields
in particle systems}, (MIT Cambridge, MA, 1987)

\bibitem{Castro03} A. Castro, A. Rubio, M.~J. Stott,
Can. J. Phys. {\bf 81}, 1 (2003)

\bibitem{fmm-perio} K.~N. Kudin, G.~E Scuseria, J. chem. Phys. {\bf 121}, 2886 (2004).

\bibitem{octopus} M.~A.~L. Marques, A. Castro, G.~F. Bertsch, A. Rubio,
Comput. Phys. Commun. {\bf 151}, 60 (2003)

\bibitem{Onida95} G. Onida, L. Reining, R.~W. Godby, R. Del~Sole and W. Andreoni, Phys. Rev. Lett. {\bf 75}, 818 (1995).

\bibitem{Jackson} J.~D. Jackson, {\it Classical Electrodynamics}, 
Wiley (1999)

\bibitem{Spataru04b} C. Spataru, S. Ismail-Beigi, L.~X. Benedict, S.~G. Louie
Applied Physics {\bf A 78}, 1129 (2004)

\bibitem{Gradshteyn} I.~S. Gradshteyn, M. Rhysik,
{\sl Tables of Integrals, Series and Products} (Academic, New York, 1980)


\bibitem{self} http://people.roma2.infn.it/$\sim$marini/self/

\bibitem{abinit} http://www.abinit.org

\bibitem{trouiller} N. Troullier, J.~L. Martins, Phys. Rev. {\bf B 43}, 1993 (1991)

\bibitem{pz} J.~P. Perdew and A. Zunger, Phys. Rev. {\bf B 23}, 5048 (1981)

\bibitem{bse} S.Albrecht, L. Reining, R. Del~Sole and G. Onida, Phys. Rev. Lett. {\bf 80}, 4510 (1998); L.~X.~Benedict, E.L. Shirley,  R.~B. Bohn, {\it ibid.} {\bf 80}, 4514 (1998); M. Rohlfing and S.G. Louie, {\it ibid.} {\bf 81 }, 2312 (1998); Pys. Rev. {\bf B 62}, 4927 (2000)

\bibitem{adler} S.~L. Adler, Phys. Rev. {\bf 126}, 413 (1962); N. Wiser, Phys. Rev. {\bf 129}, 62 (1963) 

\bibitem{Spataru04} C.D. Spataru, S. Ismail-Beigi, L.~X. Benedict, S.~G. Louie
Phys. Rev. Lett. {\bf 92}, 077402 (2004)

\bibitem{plasmonpole} R.~W. Godby, R.~J. Needs, Phys. Rev. Lett. {\bf 62}, 1169 (1989). To get converged inverse dielectric matrix  we have used unoccupied bands up to 28 eV in the electron-hole  energies, and a number of $G$ vectors until $G^2/2 = 50 eV$ in the inversion of the matrix.

\bibitem{Hybertsen} M.~S. Hybertsen and S.~G. Louie, Phys. Rev. Lett. {\bf 55}, 1418 (1985); R.W. Godby, M. Schluter, L. J. Sham, Phys. Rev. B {\bf 37}, 10159 (1988)

\bibitem{delerue} C. Delerue, G. Allan,  M. Lannoo, Phys. Rev. Lett. {\bf 90}, 076803 (2003)

\bibitem{daniele} D. Varsano, A. Marini, A. Rubio (work in progress)



\end{thebibliography}
\end{document}